\documentclass[aps,prd,nofootinbib,showkeys,floatfix,twocolumn]{revtex4-1}
\pdfoutput=1
\usepackage{amsmath}
\usepackage{amssymb}
\usepackage{graphicx}
\usepackage[mathscr]{eucal}
\usepackage{color}
\usepackage[T1]{fontenc} 
\usepackage{slashed}
\usepackage{xspace}
\usepackage{hyperref}
\usepackage{cleveref}
\usepackage{xspace}
\usepackage{ulem}
\usepackage{cancel}
\usepackage{enumerate}
\usepackage{units}
\usepackage[utf8]{inputenc}

\newcommand\SARAH{{\tt SARAH}\xspace}
\newcommand\SPheno{{\tt SPheno}\xspace}

\newcommand{\hc}{\text{h.c.}}

\newcommand{\GeV}{\text{GeV}\xspace}
\newcommand{\TeV}{\text{TeV}\xspace}
\newcommand{\DRbar}{{\ensuremath{\overline{\mathrm{DR}}}}\xspace}
\newcommand{\DRbarp}{\ensuremath{\DRbar'}}

\newcommand{\abs}[1]{\left| #1 \right|}

\newcommand{\AddrBonn}{%
Bethe Center for Theoretical Physics \& Physikalisches Institut der 
Universit\"at Bonn, \\
 53115 Bonn, Germany
}

\newcommand{\AddrCERN}{%
Theory Division, CERN, 1211 Geneva 23, Switzerland
}

\newcommand{\AddrParis}{%
1-- Sorbonne Universit\'es, UPMC Univ Paris 06, UMR 7589, LPTHE, F-75005, Paris, France \\
2-- CNRS, UMR 7589, LPTHE, F-75005, Paris, France}

\graphicspath{{./Figs/}}

\begin{document}

\hfill BONN--TH--2015--11, CERN-PH-TH-2015-260 \vspace{0.2cm}

 \title{The Higgs Mass in the MSSM  at two-loop order\\ beyond minimal flavour violation}

\author{Mark D. Goodsell}
\email{goodsell@lpthe.jussieu.fr}
\affiliation{\AddrParis}

\author{Kilian Nickel} 
\email{nickel@th.physik.uni--bonn.de}
\affiliation{\AddrBonn}
 
 \author{Florian Staub}
 \email{florian.staub@cern.ch}
 \affiliation{\AddrCERN}


\begin{abstract}
Soft supersymmetry-breaking terms provide a wealth of new potential sources of flavour violation, which are tightly constrained by precision experiments. This has posed a challenge to construct flavour models which both explain the structure of the Standard Model Yukawa couplings and also predict soft-breaking patterns that are compatible with these constraints. While such models have been studied in great detail, the impact of flavour violating soft terms on the Higgs mass at the two-loop level has been assumed to be small or negligible. In this letter, we show that large flavour violation in the up-squark sector can give a positive or negative mass shift to the SM-like Higgs of several GeV, without being in conflict with other observations. 
We investigate in which regions of the parameter space these effects can be expected. 
\end{abstract}
\maketitle

\section{Introduction}
\label{sec:intro} 
The discovery of the Higgs boson \cite{Chatrchyan:2012ufa,Aad:2012tfa} has already celebrated its 
third anniversary, and in the meantime its properties are measured with an impressive precision. 
The average mass is \mbox{$m_h=(125.09 \pm 0.32)~\GeV$} \cite{Aad:2015zhl,Khachatryan:2014jba,Aad:2014aba}. 
This measurement is much better than the theoretical prediction of the mass in any model beyond the standard model (SM). 
The most studied 
extension of the SM is the minimal supersymmetric standard model (MSSM), in which the uncertainty 
is estimated to be of the order of a few GeV, taking into account the dominant two-loop corrections \cite{Degrassi:2002fi,Heinemeyer:2004gx,Buchmueller:2013psa,Hahn:2013ria}. 
However, this estimate does not include the impact of large flavour violation for instance.
While one-loop contributions to the Higgs mass are known exactly, the widely used  
two-loop corrections (those not involving electroweak gauge couplings) make the approximation of including only third generation states. 
Electroweak corrections have been calculated for the MSSM at $\mathcal O(\alpha \alpha_s)$ \cite{Degrassi:2014pfa}, stemming from the D-term couplings between Higgs bosons and coloured sfermions and neglecting all fermion masses other than the top mass.
\footnote{In \cite{Degrassi:2014pfa}, corrections of $\mathcal O(\alpha \alpha_s)$ are calculated together with momentum-dependent $\mathcal{O}(\alpha_t\alpha_s)$ corrections because both are expected to be suppressed by $\mathcal{O}(m_Z^2/m_t^2)$ compared to $\mathcal{O}(\alpha_t\alpha_s)_{p^2=0}$.
Interestingly, since the Higgs mass is of the order of the electroweak scale, those contributions required the inclusion of all $\mathcal{O}(\alpha \alpha_s)$ terms -- and the first two generations via their D-term coupling only. However, those corrections are almost completely orthogonal to the (much larger) contributions considered here -- and indeed in that calculation flavour mixing was neglected.}\\
The impact of first and second generation (s)quarks (in gaugeless contributions) can safely be neglected under the assumption 
that the only source of flavour violation is the CKM matrix of the SM. In this ansatz, known 
as 'minimal flavour violation' \cite{Hall:1985dx,D'Ambrosio:2002ex,Cirigliano:2005ck}, 
the three generations of sfermions are aligned with the corresponding 
fermions and the soft-breaking terms do not introduce any additional flavour violation. 
However, there is no fundamental reason why this alignment should be present. 
In particular in models where SUSY breaking is transmitted via gravity, this is often a very strong and hard to motivate assumption \cite{Kallosh:1995hi,Kribs:2010md}. 
It can be motivated in models with pure gauge mediation, but these models have significant difficulties in explaining the Higgs 
mass -- hence recent interest in non-minimal gauge mediation models with direct couplings between the messenger and visible sectors, which may as a consequence lead to flavour violation \cite{Shadmi:2011hs,Calibbi:2013mka,Galon:2013jba,Brummer:2013upa,Abel:2014fka,Delgado:2015bwa}. \\
For these reasons, non-minimal flavour violation in the MSSM has been studied for several years: 
the focus has been mainly on the collider phenomenology \cite{delAguila:2008iz,Bartl:2010du,Bartl:2012tx,Blanke:2013zxo,Backovic:2015rwa,DeCausmaecker:2015yca},
and the impact on flavour precision observables, see for instance Ref.~\cite{Buchalla:2008jp} and references therein. 
Interestingly, a large mixing between stops and scharms could explain some recent flavour anomalies \cite{Altmannshofer:2014rta}. 
It is also known that large flavour mixing involving stops can have an important consequence for the Higgs 
mass calculation at one-loop \cite{Heinemeyer:2004by,Heinemeyer:2004gx,Cao:2007dk,Kowalska:2014opa,Brignole:2015kva,AranaCatania:2011ak,Arana-Catania:2014ooa}. A complete one-loop calculation including all flavour and momentum effects exists for years for the MSSM. Already at one-loop, it has been shown \cite{AranaCatania:2011ak,Arana-Catania:2014ooa} that Higgs mass corrections can be larger than 10~\GeV{} due to large flavour mixing in the squark sector, and that positive as well as negative differences can be found. The shifts can even be as high as $\mathcal{O}(60)~\GeV{}$ if more mixing parameters are included. However, those points are highly constrained by precision $B$ observables, namely $B_s\to \mu^+\mu^-$, $B\to X_s \gamma$ and $\Delta M_{B_s}$.
Knowing that these effects can be large at one loop, it was not yet studied how significant these effects can be at two loops. 
We close this gap here. We shall show that the usually neglected two-loop corrections can shift the 
Higgs mass by several GeV. \\
This letter is organised as follows: in sec.~\ref{sec:model} we introduce our conventions to parametrise 
flavour violation in the MSSM, before we show the numerical results in sec.~\ref{sec:numerics}. 
We discuss the results 
in sec.~\ref{sec:discussion}.

\section{The MSSM with general flavour violation}
\label{sec:model}
We shortly introduce our conventions for the discussion in the following. We stick closely to the 
SLHA 2 conventions for the definition of our basis \cite{Allanach:2008qq}, and the superpotential reads
\begin{align}
W = Y^{i j}_{e}\,\hat L_i \hat E_j \hat H_d 
   +Y^{i j}_{d}\, \hat Q_i \hat D_j \hat H_d 
   +    Y^{i j}_{u} \hat Q_i \hat U_j \hat H_u +
 \mu\, \hat H_u \hat H_d 
\end{align}
where the sums over colour and isospin indices are implicit. 
The hat symbol (e.g. $\hat L_i$) denotes a superfield. 
In general, the 
Yukawa couplings $Y_X$ ($X=e,d,u$) are $3\times 3$ complex matrices. 
Since there is no source of 
lepton flavour violation in the MSSM, $Y_e$ has to be diagonal: $Y_e=\text{diag}(y_e,y_\mu,y_\tau)$. 
Moreover, one can always perform a rotation into the super-CKM basis 
where quark Yukawa couplings become diagonal as well:
\begin{equation}
  \label{eq:yukawadiagonal}
  Y_d=\text{diag}(y_d,y_s,y_b),\quad Y_u=\text{diag}(y_u,y_c,y_t)\quad .
\end{equation}
The entire information of flavour violation is then included in the CKM matrix $V$ which is defined as 
\begin{equation}
V = (U_L^u)^{\dagger} U_L^d
\end{equation}
where $(U_L^d)$ and $(U_L^u)$ rotate the left-handed down- and up-quarks which are assumed to be aligned 
with the corresponding superfields. \\
The soft-SUSY breaking sector of the model is parametrised by
\begin{align}
 - \mathscr{L} =  &  \big(T^{ij}_e \tilde{l}_i \tilde{e}_j H_d + T^{ij}_d \tilde{q}_i \tilde{d}_j H_d+T^{ij}_u \tilde{q}_i \tilde{u}_j H_u \nonumber \\ & 
 \hspace{1cm} + B_\mu H_d H_u   + \hc \big) \nonumber \\
 & + \left(M_1 \lambda_B \lambda_B + M_2 \lambda_W \lambda_W + M_3 \lambda_G \lambda_G  + \hc \right)  \nonumber \\
 &  +  m^2_{\phi,{ij}} \tilde{\phi}^*_i \tilde{\phi}_j  + m_{H_d}^2 |H_d|^2 + m_{H_u}^2 |H_u|^2  
\end{align}
with $\phi=u,d,q,e,l$. In the limit of minimal flavour violation, $T_i = A_i Y_i$ ($i=d,u,e$) would hold, but we want 
to study explicit deviations from this. We concentrate in the following on the up-squark sector. 
In general, the mass matrix squared for the six up-squarks, $M_U^2$, in the basis $(\tilde u_L, \tilde c_L, \tilde t_L, \tilde u_R, \tilde c_R, \tilde t_R)$ is given by
\begin{align}
  M^2_{U} &=
  \begin{pmatrix}
    V m^2_{q} V^\dagger+\frac 12 v_u^2 Y_{u}^2 + D_{LL} & X^\dagger \\
    X & m^2_u + \frac 12 v_u^2 Y_{u}^2  + D_{RR}
  \end{pmatrix}
\end{align}
with the $3\times 3$ matrix
\begin{equation}
X = - \frac{v_d}{\sqrt 2} \mu^* Y_u + \frac{v_u}{\sqrt 2} T_u
\end{equation}
and the $D$-term contributions are 
expressed in diagonal matrices $D_{LL}$ and $D_{RR}$. We assume further that the only sources of additional flavour violation 
are the (2,3) and (3,2) entries of $T_u$ and that $T_d$ as well as $T_{u,11}$ are vanishing. 
In this case, we can parametrise the squark sector by:
\begin{eqnarray*}
&m_{u,33}, \ m_{u,22}, \ m_{q,33}, \ m_{q,22} & \\ 
& \tilde m \equiv m_{q,11} = m_{u,11} = m_{d,ii}& \\
&\ T_{u,33},\ T_{u,32},\ T_{u,23}& \\
&\mu,\, \tan\beta&
\end{eqnarray*}
For simplicity, we assumed a universal mass $\tilde m$ for all squarks not mixing with the stops, and take this value also for all slepton soft masses. The remaining parameter is the gluino mass $M_3$, which will be important in the following. 

\section{Numerical results}
\label{sec:numerics}
For the numerical analysis we make use of the combination of the public codes \SPheno \cite{Porod:2011nf,Porod:2003um} and \SARAH \cite{Staub:2009bi,Staub:2010jh,Staub:2011dp,Staub:2012pb,Staub:2013tta,Staub:2015kfa}. All masses are renormalised in the 
$\DRbarp$ scheme: at one-loop all corrections including the full momentum dependence are taken into account for any SUSY and Higgs state. 
For the neutral Higgs masses, the two-loop corrections in the gaugeless limit and without momentum dependence are included, but all generations of (s)fermions are taken into account \cite{Goodsell:2014bna,Goodsell:2015ira}.
These results make use of the generic approach developed in Refs.~\cite{Martin:2001vx,Martin:2003it,Martin:2005eg}, which also contain a detailed description of the renormalisation procedure for the interested reader.
We will refer to this calculation as $m_h^{\text{full}}$ in the following, keeping in mind that these provisos exist. It has been shown that the obtained results for the MSSM are in perfect agreement with widely used 
results of Refs.~\cite{Brignole:2001jy,Degrassi:2001yf,Brignole:2002bz,Dedes:2002dy,Dedes:2003km}, if first and second generation (S)quarks are neglected  (this will be called $m_h^{\text{approx}}$ in the following). If they are taken into account in the 
limit of minimal flavour violation, the differences are still very small. We shall study what happens if we are far away from minimal flavour violation. 

\subsection{Exploring the MSSM with large stop flavour violation}
\label{sec:exploring}
\begin{figure}[hbt]
\includegraphics[width=\linewidth]{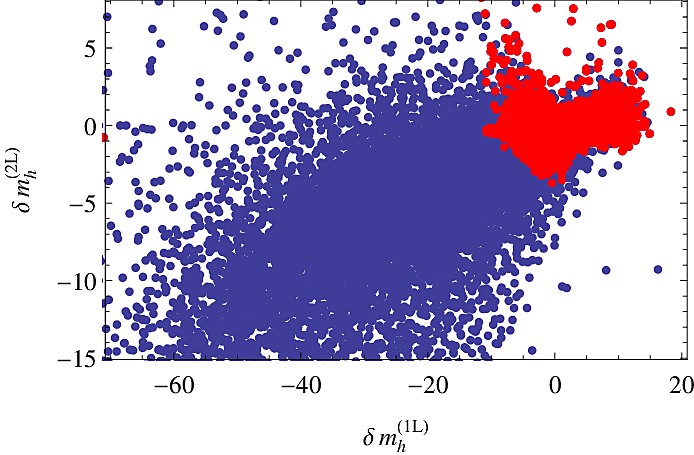}
\caption{Correlation between $\delta m_h^{(1L)}$ and $\delta m_h^{(2L)}\equiv \delta m_h$. The blue points are all points which give a tachyon-free spectrum without any further restrictions. The red points provide at two loops a Higgs mass with $m_h > 120$~GeV.}
\label{fig:1L2L}
\end{figure}

We fix in the following the parameters which have a less important impact on the two-loop Higgs mass corrections as follows:
\begin{eqnarray*}
& M_1 = 100~\GeV,\,M_2=200~\GeV,\, \tilde m = 1500~\GeV& \\
&\mu=500~\GeV,\,M_A^2 = (1000~\GeV)^2,\,\tan\beta=10&
\end{eqnarray*}
For the other parameters, we scan over the following ranges:
\begin{eqnarray*}
&M_3 \in [1,3]~\TeV & \\
&m_{u/q,33} \in [0.2,2]~\TeV,\, m_{u/q,22} \in [1.2,2.5]~\TeV&\\
&T_{u,ij} \in [-4,4]~\TeV \ (i,j=2,3)&
\end{eqnarray*}
To be consistent with LHC collider limits, the second generation mass parameters are chosen larger than $\unit[1.2]{\TeV}$. 
The third generation can be much lighter. It is always possible to choose $M_1$ such that the LSP mass is close to the stop mass, thus avoiding the collider limits. The change of $M_1$ would have no impact on the results. 
The quantity of interest is the difference between the calculation with third generation squarks only ($m_h^{\rm approx}$) and the 'full' calculation including all generations, $m_h^{\rm full}$,\footnote{The specific settings in \SPheno used for the two values of the Higgs mass are in the Flag 8 of {\tt Block SPhenoInput}: the value is set to 3 for $m_h^{\rm full}$ (diagrammatic calculation) and to 9 for $m_h^{\rm approx}$ (2-loop dominant, 3rd generation contributions; using routines based on Refs.~\cite{Brignole:2001jy,Degrassi:2001yf,Brignole:2002bz,Dedes:2002dy,Dedes:2003km}). }  
\begin{align}
\delta m_h^{(2L)}\equiv \delta m_h = m_h^{\rm full} - m_h^{\rm approx},
\end{align}
If we impose no cut upon the Higgs mass -- i.e. do not require it to have the observed value of $125$ GeV -- then we can have very large shifts in its value through flavour effects.
To begin with, we consider a rough scan over 250k points, where the only requirement is that the spectrum contains no tachyons, leaving 95k points. Those are shown in 
 Fig.~\ref{fig:1L2L}, where $\delta m_h^{(1L)}$, the difference between a full one-loop calculations and the one-loop calculation neglecting flavour effects, is shown against $\delta m_h^{(2L)}$. 
We see that there is indeed a weak correlation between the one- and two-loop effects. The more realistic points with the restriction $m_h>\unit[120]{\GeV}$ are shown in red in Fig.~\ref{fig:1L2L}. Cutting on points which have a sufficiently large Higgs mass at the two-loop level singles out points 
where the one- and two-loop effects are of comparable size. 

\begin{figure}[hbt]
\includegraphics[width=0.82\linewidth]{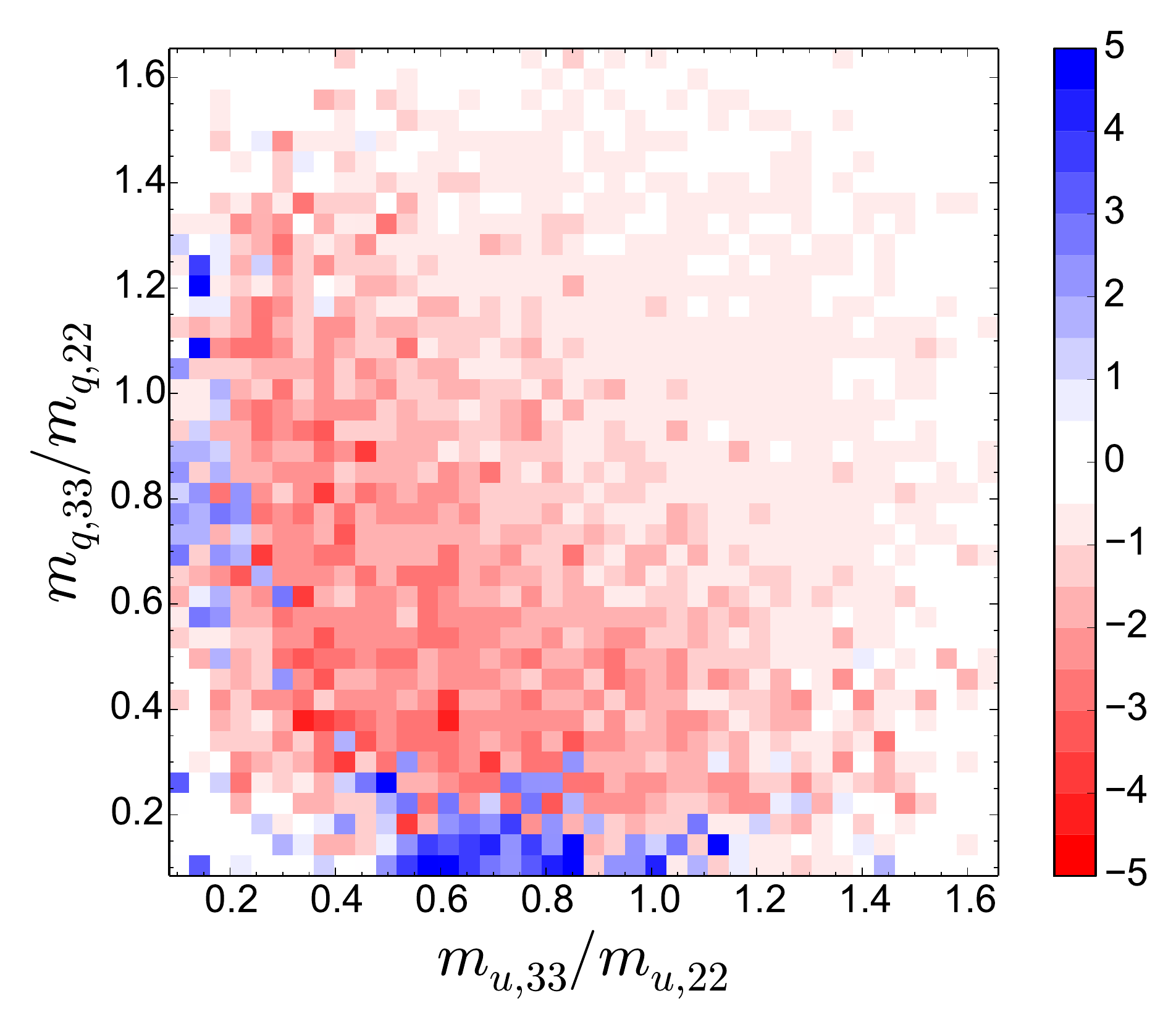} \\
\includegraphics[width=0.82\linewidth]{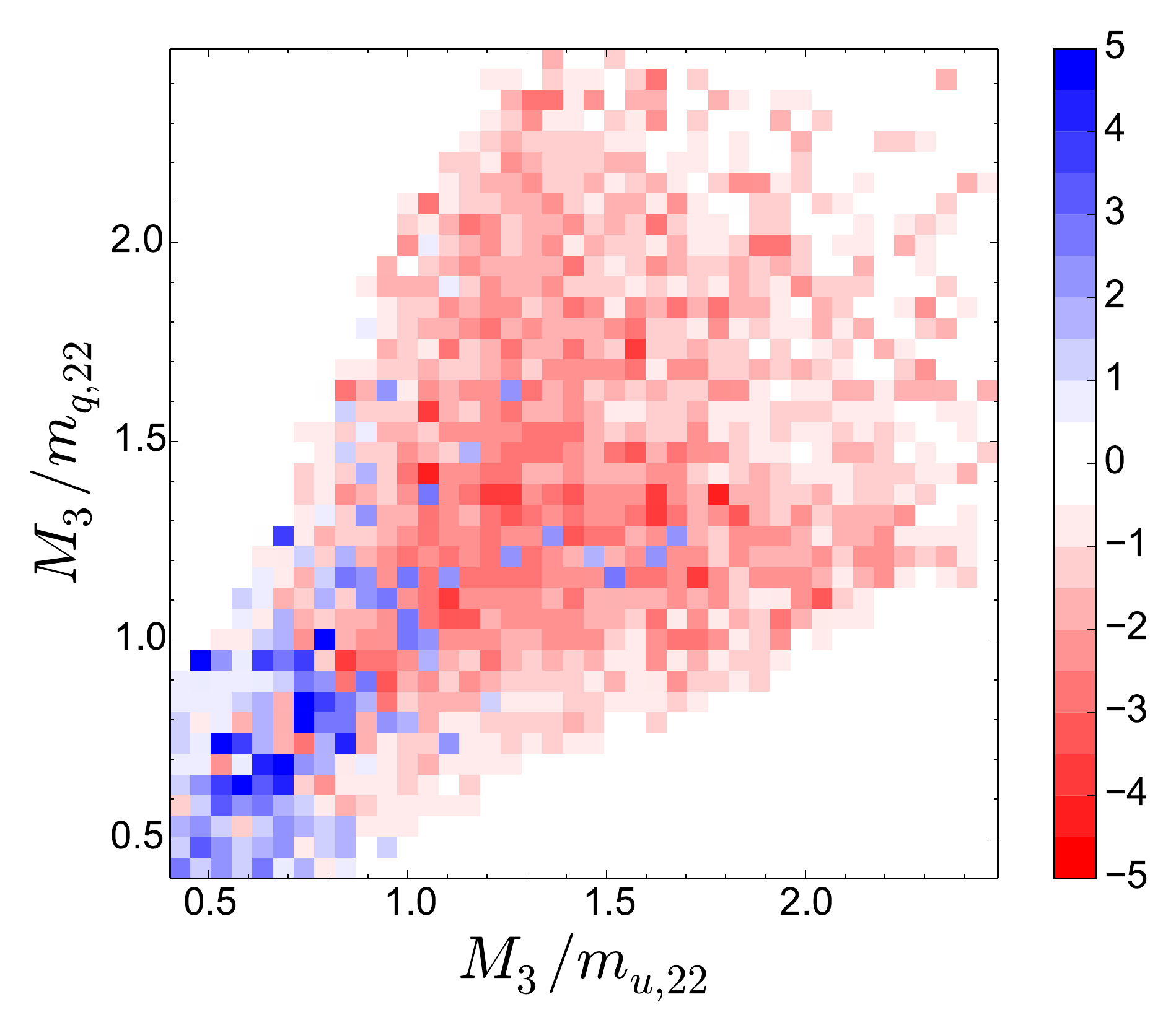}\\
\includegraphics[width=0.82\linewidth]{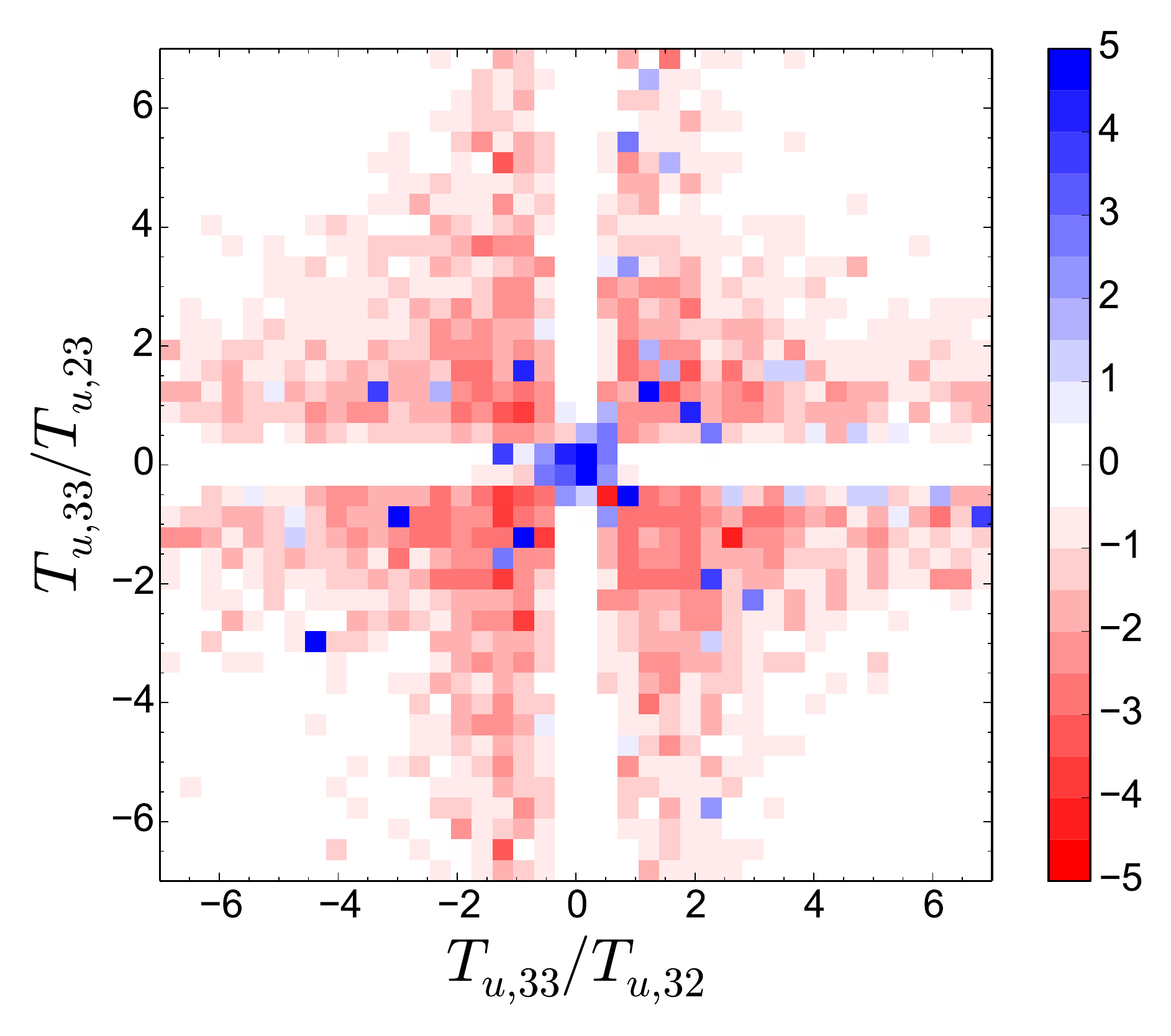}
\caption{$\delta m_h$ (in GeV) of the point with the maximal $|\delta m_h|$ per bin is shown, as function of different ratios of important soft-breaking parameters.}
\label{fig:Density}
\end{figure}

\begin{figure}[hbt]
  \centering
  \includegraphics[width=0.78\linewidth]{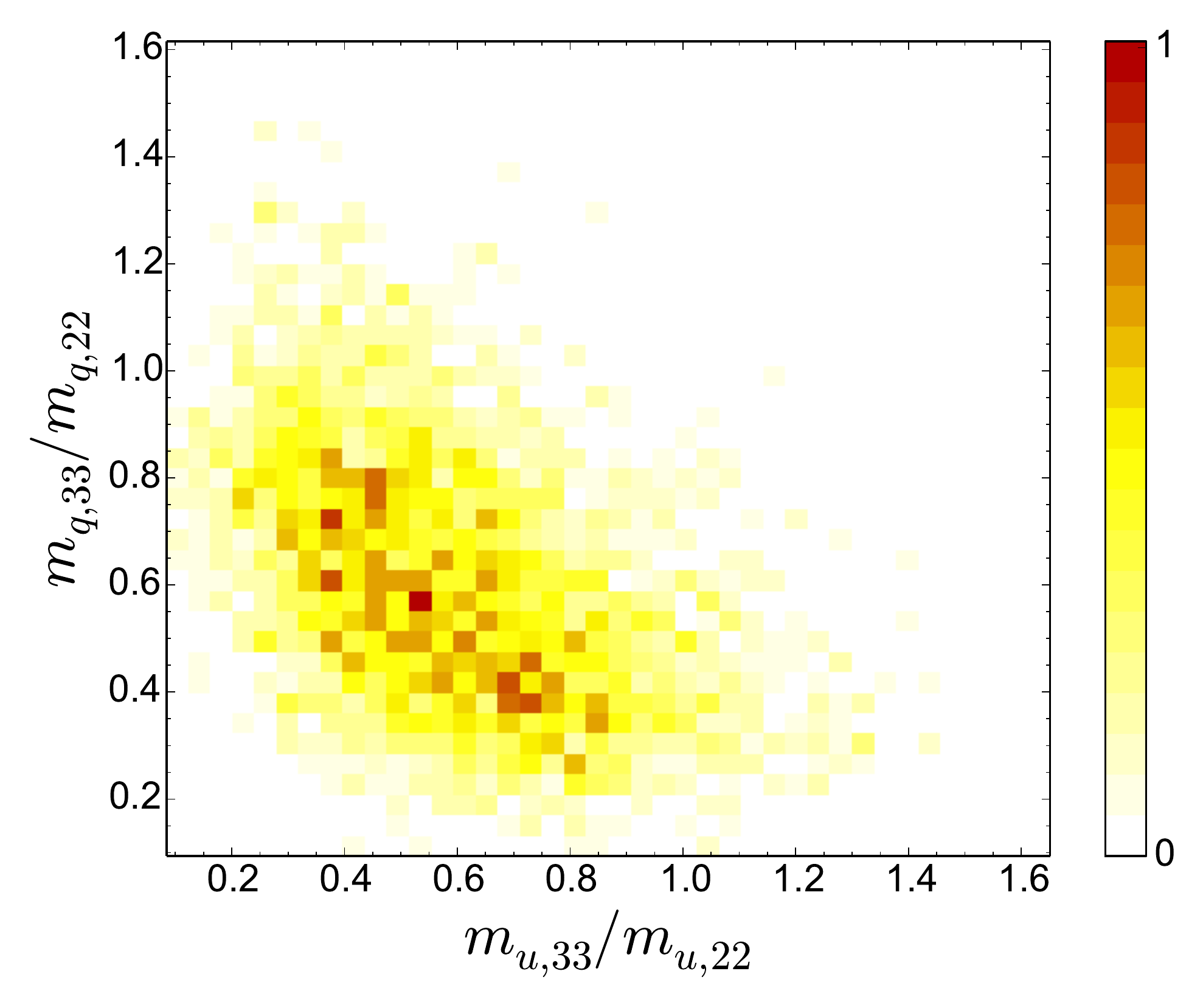} \\[5mm]
  \includegraphics[width=0.78\linewidth]{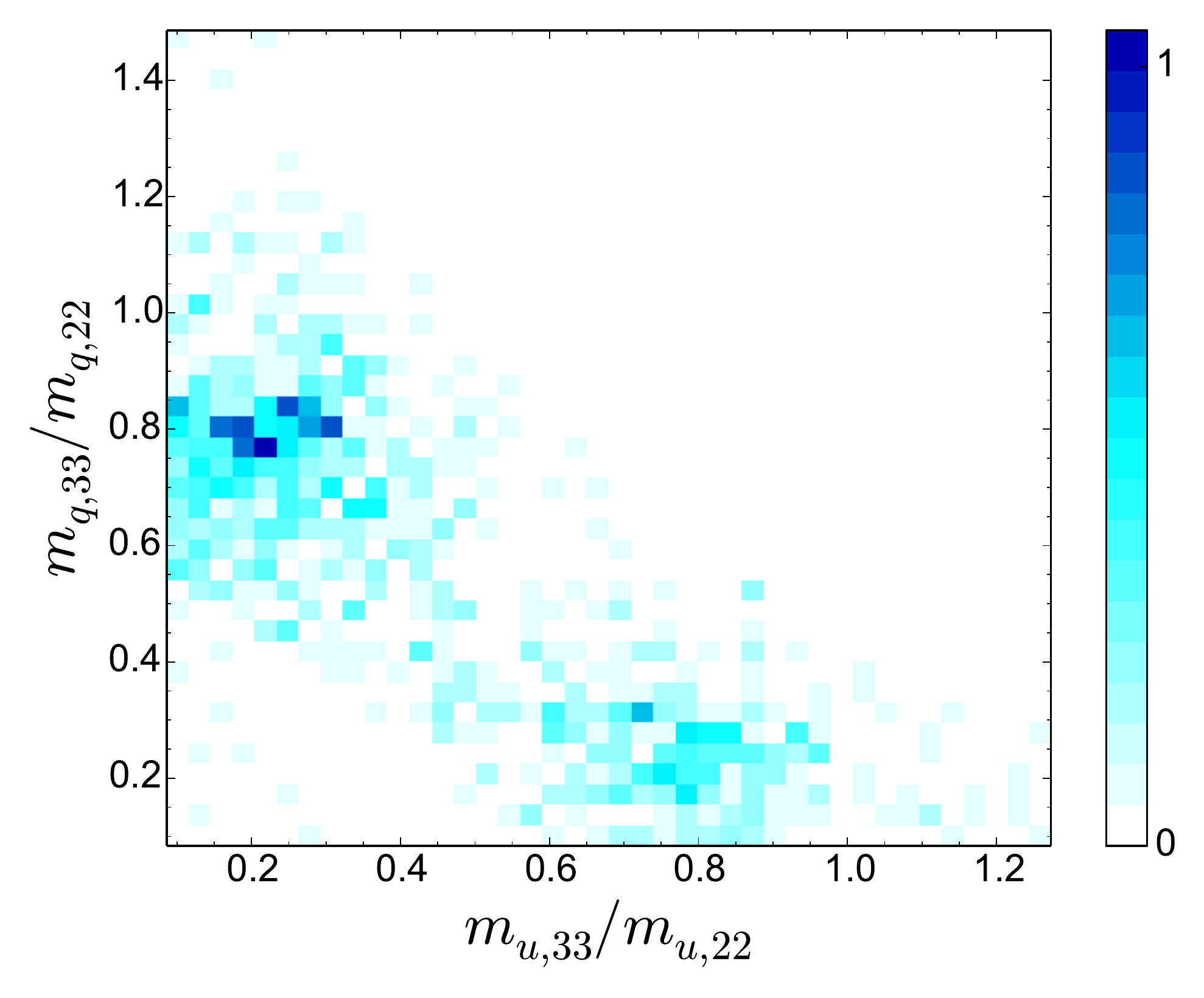} 
  \caption{These plots show normalised histograms of points from the generated sample that fulfil $|\delta m_h|\in[0.8, 7]$ \GeV{} (color bars range from 0 to 1). The upper plot shows points with negative $\delta m_h$ (red hue) and the lower plot shows points of positive $\delta m_h$ (blue hue).}
  \label{fig:hist2d}
\end{figure}

\begin{figure}[hbt]
\includegraphics[width=\linewidth]{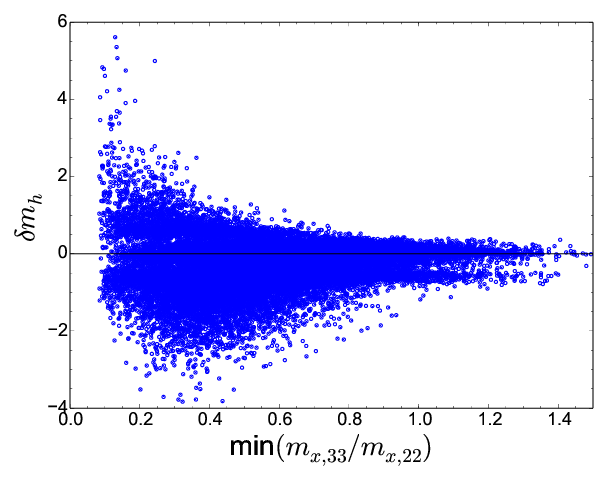}
\caption{$\delta m_h$ as function of $\text{min}(m_{x,33}/m_{x,22})$ ($x=q,u$).}
\label{fig:SoftRatio2D}
\end{figure}

\begin{figure}[hbt]
\includegraphics[width=\linewidth]{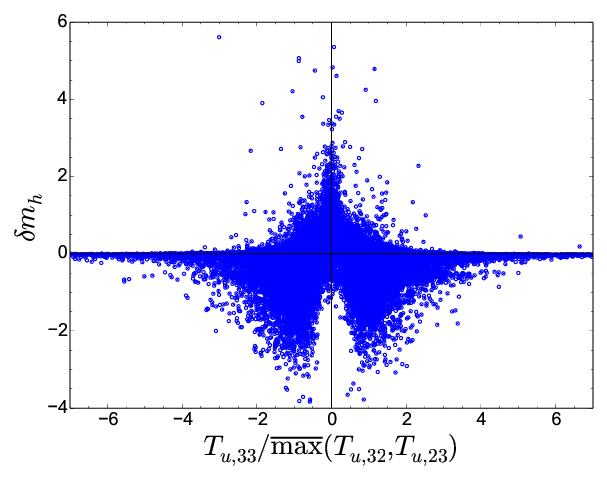}
\caption{$\delta m_h$ as function of $T_{u,33}/\overline{\text{max}}(T_{u,32},T_{u,23})$, where $\overline{\text{max}}$ 
picks the entry whose absolute value is larger independent of the sign.}
\label{fig:Trilinears}
\end{figure}

To investigate further,  we shall be interested in potentially phenomenologically relevant models, and so shall show the results of a larger, finer scan where we restrict to  points for which the full Higgs mass including all 
generation of squarks at two loops ($m_h^{\rm full}$) is larger than 120~GeV. 
This scan included 5 million points using a flat prior. To avoid the issue of undersampling in a scan with six free parameters, at least $10^6$ points have to be sampled, which is far exceeded by this number. From the total number of points, a selection of about 50k points have $m_h>\unit[120]{\GeV}$ and $\abs{\delta m_h}> \unit[0.5]{\GeV}$, as well as fulfilling flavour constraints from all important $B$ observables. The strongest constraints usually come from $b\to s\gamma$. This selection is used throughout the following plots, Figs.~\ref{fig:Density}-\ref{fig:Trilinears}.
It is useful to define the ratio $r_u\equiv m_{u,33}/m_{u,22}$ and $r_q\equiv m_{q,33}/m_{q,22}$ of soft mass parameters.
We show in Fig.~\ref{fig:Density} the value of $\delta m_h$ with the 
largest absolute value per bin. These plots indicate regions where the {\it highest} corrections, positive as well as negative, can be obtained, possibly among other points with smaller corrections residing in the same bin which are not shown. Therefore, each plot in Fig.~\ref{fig:Density} projects out a certain amount of points and the remaining number equals the number of bins.
Complementarily, Fig.~\ref{fig:hist2d} shows histograms of the number of points (arbitrary units) which survive the cut $|\delta m_h| \in \unit[\lbrack 0.8, 7\rbrack]{\GeV}$. The upper plot (red hue) shows only points with negative $\delta m_h$ and the lower plot (blue hue) shows only positive $\delta m_h$. 
These plots do not show the magnitude of the corrections, but rather the general location in parameter space where positive and negative corrections can be found.
We find the following behaviour:
\begin{enumerate}[(i)]
 \item From Fig.~\ref{fig:Density} (upper) it can be seen that a necessary condition for a large deficit of several GeV in $m_h^{\rm approx}$  (i.e. $\delta m_h>0$) is a large hierarchy between the third and second generation of the soft-masses $m_q$ or $m_u$ (i.e. small values of $r_u,r_q$).  In particular, many such points reside in a region around $(r_u,r_q)=(0.8, 0.2)$ and $(0.8, 0.2)$ which is visible in Fig.~\ref{fig:hist2d} (lower).
On the other side, if $r_q$ or $r_u$ is $\geq$ 0.4, one finds negative $\delta m_h$. This can be seen in Fig.~\ref{fig:hist2d} (upper), where the bulk of points is within the area of $r_u\geq 0.4, r_q \geq 0.4$. It is also visible in Fig.~\ref{fig:SoftRatio2D}, where $\delta m_h$ is displayed against $\text{min}(r_u, r_q)$. Large negative values of $\delta m_h$ are found around 0.4.
 \item In the case that the gluino is lighter than the second generation of soft-masses ($M_3/m_{x,22}<1,\,x=q,u$), $\delta m_h$ is found positive (blue area within Fig.~\ref{fig:hist2d}, middle), while for a heavier gluino ($M_3/m_{x,22}>1,\,x=q,u$) the additional corrections from  flavour violation are negative (red area within Fig.~\ref{fig:hist2d}, middle).
 \item The sign of the additional corrections depends strongly on the ratio of $T_{u,33}$ and the two off-diagonal couplings $T_{u,32}$ and $T_{u,23}$. If $|T_{u,32}|$ or $|T_{u,23}|$ are much bigger than $|T_{u,33}|$, the flavoured two-loop 
 corrections are usually large and positive. Negative corrections appear in particular for the case that 
 $\text{max}(|T_{u,32}|,|T_{u,23}|) \simeq |T_{u,33}|$. 
This is shown in Fig.~\ref{fig:Density} (lower) and also in Fig.~\ref{fig:Trilinears}.
We checked that a very similar pattern as in Fig.~\ref{fig:Trilinears} also exists at one loop: positive (negative) corrections can be found around $T_{u,33}/\overline{\text{max}}(T_{u,32},T_{u,23})=0$ (at $\pm 1$, respectively), whereas the magnitude can be much larger.
\end{enumerate}

\subsection{Examples}
\label{sec:benchmark}
To further investigate the dependence on the different parameters, we pick two parameter points 
where the flavour effects at two-loop give either a positive or negative shift to the Higgs mass. 

\subsubsection{Positive contributions  from flavour effects}
\begin{figure}[hbt]
\begin{center}
\includegraphics[width=0.75\linewidth]{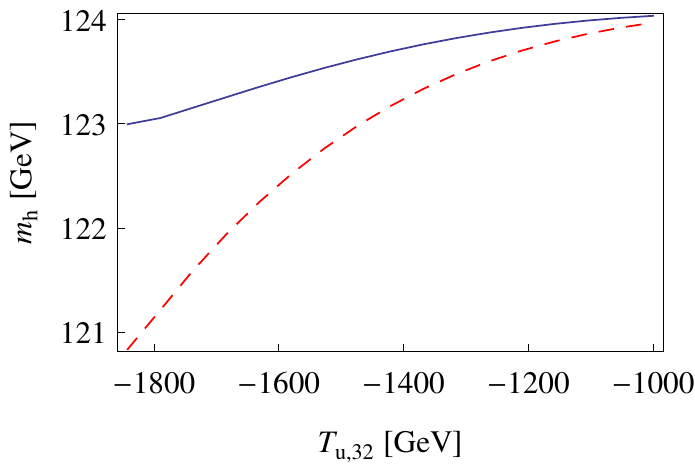}  \\
\includegraphics[width=0.75\linewidth]{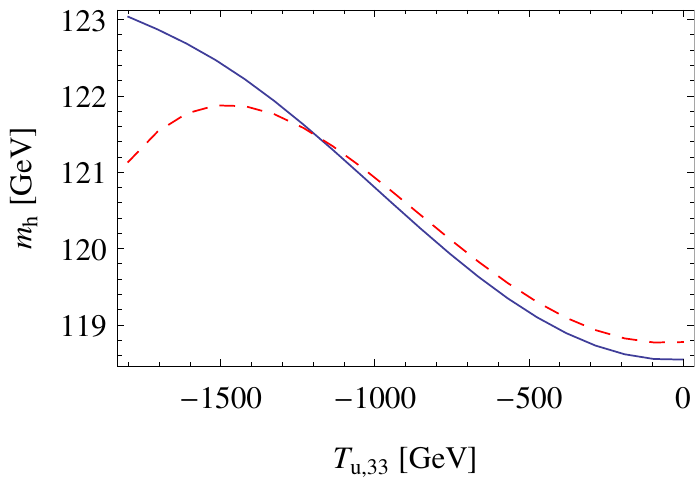}  \\
\includegraphics[width=0.75\linewidth]{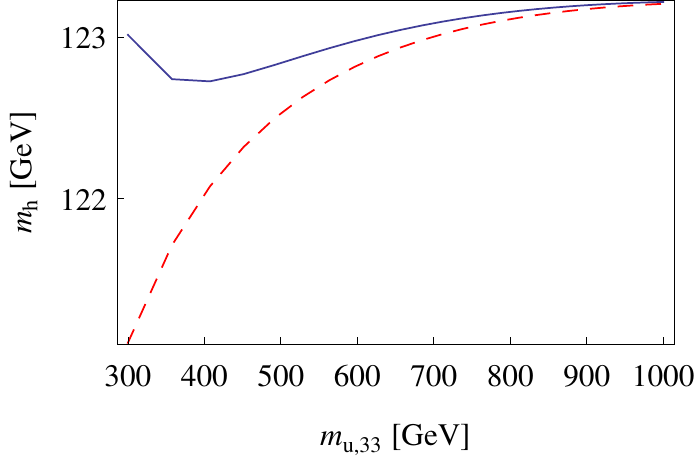} \\
\includegraphics[width=0.75\linewidth]{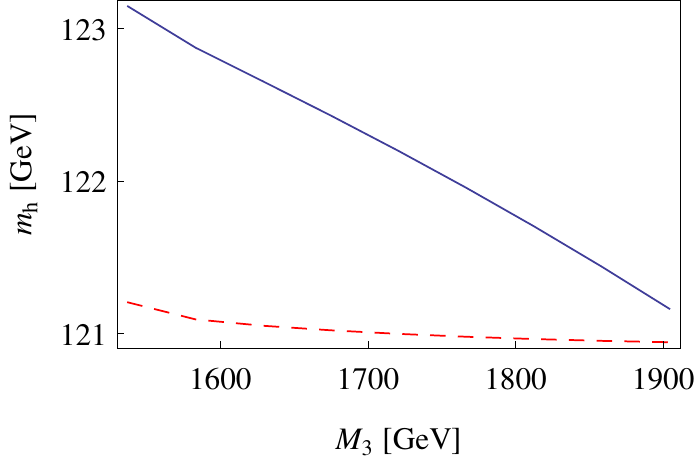} 
\end{center}
\caption{$m_h^{\text{full}}$ (solid blue) and $m_h^{\text{approx}}$ (dashed red) as functions of $T_{u,32}$, $T_{u,33}$, $m_{u,33}$ and $M_3$. The other parameters are fixed 
to the values in eq.~(\ref{eq:BP}). }
\label{fig:BP}
\end{figure}

The input parameters of the first example are
\begin{eqnarray}
 &m_{u,33} = 300~\GeV ,\, m_{q,33} = 2000~\GeV & \nonumber \\
 &m_{u,22} = m_{q,22} = 2300~\GeV & \nonumber \\
 &T_{u,33} =  T_{u,32} = -1800~\GeV ,\, T_{u,23} = 0,\, & \nonumber  \\
 \label{eq:BP}
 & M_3 = 1550~\GeV.\, &
\end{eqnarray}
Note that this choice of parameters respects direct collider bounds by the same reasoning that was given in sec.~\ref{sec:exploring}.
Depending on the used two-loop calculation, we find the following values for the SM-like Higgs mass
\begin{align}
m_h^{\rm full} = 123.1~\GeV& \\
m_h^{\rm approx} = 121.1~\GeV& 
\end{align}
Thus, the approximation to consider only the third generation (s)quark effects at two-loop gives a result 
which is 2 GeV too small compared to the full calculation. 
For comparison we checked the impact of including/excluding
the flavour violating effects at the one-loop level and found: $m_h^{\rm full,(1L)} = 116.4~\GeV$, $m_h^{\rm approx,(1L)} = 119.5~\GeV$. 
Thus, the effects are of similar size but with different sign.
The chosen point is not one of the points which maximizes the difference
between both calculations, but it can be used to see nicely the dependence on the different parameters as 
shown in Fig.~\ref{fig:BP}: the difference between both calculations quickly increases for smaller $M_3$ and 
$m_{u,33}$ as well as for larger negative $T_{u,32}$. For decreasing  $|T_{u,33}|$  the change in sign can also
be observed.\\
There is one final comment in order: it is known that large trilinear couplings in the squark sector together with a sizeable splitting in the 
soft-masses can trigger charge and colour breaking \cite{Camargo-Molina:2013sta, Blinov:2013fta,
  Chowdhury:2013dka, Camargo-Molina:2014pwa, Chattopadhyay:2014gfa}. For non-violating trilinears, the generally-used rule of thumb is that
\begin{align}
T_{u,33}^2 <& 3 |Y_u^{33}|^2 ( m_{q,33}^2 + m_{u,33}^2 + m_{H_u}^2 + |\mu|^2),
\end{align}
but once we allow new flavour-violating directions a new minimum can appear along a D-flat direction where e.g. $\langle Q_3 \rangle = \langle U_2 \rangle$, giving
\begin{align}
T_{u,23}^2<& |Y_u^{33}|^2 ( m_{q,22}^2 + m_{u,33}^2 + m_{H_u}^2 + |\mu|^2) 
\end{align}
and similarly for $2 \leftrightarrow 3$. Since the two conditions are not dramatically different, having the same scan ranges for conventional and flavour-violating trilinears is entirely reasonable and there is no reason to suspect any new problems from vacuum stability. However, to perform a careful analysis of this we checked the vacuum stability a sampling of the surviving points (after all other cuts) with {\tt Vevacious} \cite{Camargo-Molina:2013qva} allowing the possibility that the second and third generation of up-squarks can receive vacuum expectation values. We actually found that this happens 
at the global minimum of the scalar potential for the benchmark point above. However, the lifetime of this point calculated with {\tt CosmoTransitions} \cite{Wainwright:2011kj} turns out 
to be many times the age of the universe. Also all flavour observables were checked to 
be in agreement with experiment using the {\tt FlavorKit} functionality of \SARAH/\SPheno \cite{Porod:2014xia}.

\subsubsection{Negative contributions from flavour effects}

\begin{figure}[thb]
\begin{center}
\includegraphics[width=0.86\linewidth]{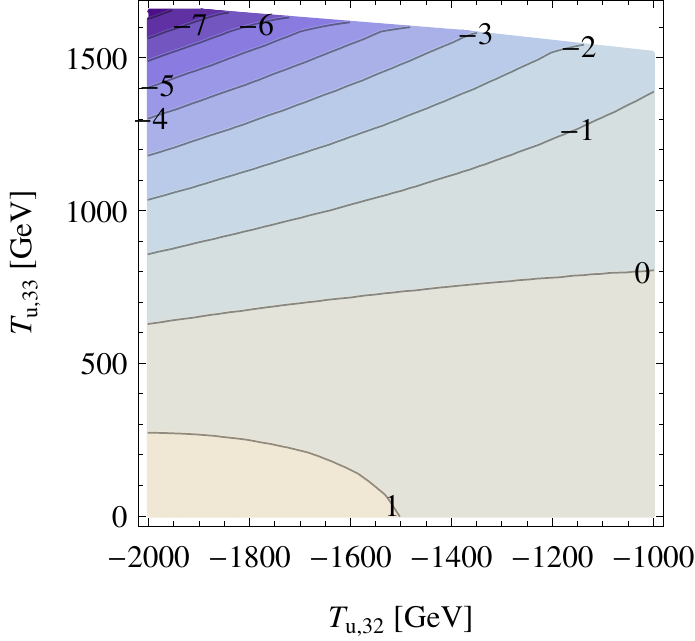}  \\
\includegraphics[width=0.85\linewidth]{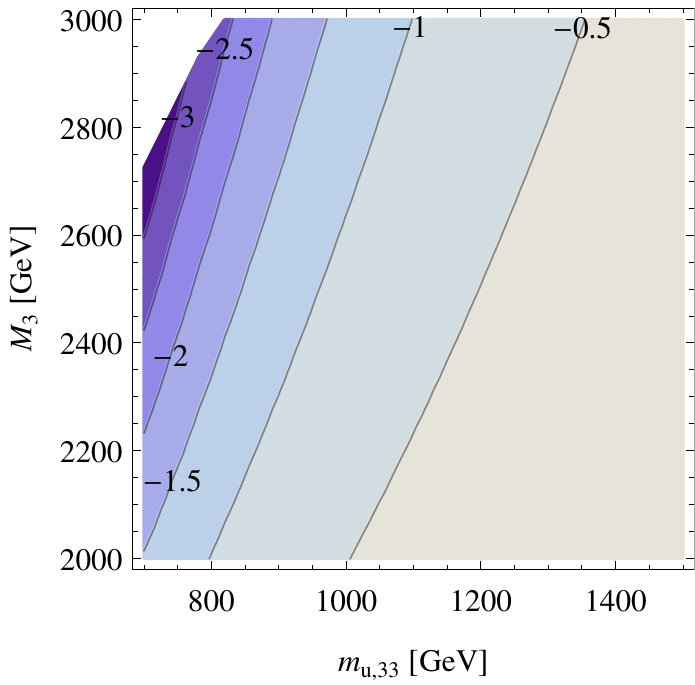}  
\end{center}
\caption{$\delta m_h$ in the $(T_{u,32},T_{u,33})$ and $(m_{u,33}, M_3)$ plane. The other parameters are fixed 
to the values in eq.~(\ref{eq:BP2}). }
\label{fig:BP2}
\end{figure}

As second example we choose the point given by 
\begin{eqnarray}
&m_{u,33} = 720~\GeV ,\, m_{q,33} = 875~\GeV & \nonumber \\
 &m_{u,22} = m_{q,22} = 2500~\GeV & \nonumber \\
 &T_{u,33} = 1200~\GeV,\, T_{u,32} = -1900~\GeV ,\, T_{u,23} = 0,\, & \nonumber  \\
 \label{eq:BP2}
 & M_3 = 2600~\GeV,\, & 
\end{eqnarray}
The scalar potential is even stable for this point and charge/colour is unbroken at the global minimum. 
The approximate calculation turns out to predict a Higgs mass which is too large by about 3~\GeV
\begin{align}
m_h^{\rm full} = 121.2~\GeV& \\
m_h^{\rm approx} = 124.0~\GeV& 
\end{align}
Just by comparing the effects from flavour violation, we already obtain an uncertainty of 3~GeV.
Thus, the  total theoretical uncertainty of $m_h$ is certainly above that widely misused estimate: the 3~GeV uncertainty stated in literature was derived for the MSSM under several assumptions like negligible effects from flavour violation. One has to be more careful in applying this uncertainty estimate to a study.
We find for this particular point that the flavour violation effects at the two-loop level are even more important than at one-loop where they cause only a shift of about 1 GeV: $m_h^{\rm full,(1L)} = 117.3~\GeV$, $m_h^{\rm approx,(1L)} = 118.3~\GeV$.
The dependence  
on the trilinear squark couplings $T_{u,32}$ and $T_{u,33}$ as well as on $m_{u,33}$ and $M_3$ is shown in Fig.~\ref{fig:BP2}.
Note, in the regions with large $|T_{u,32}|$ in the upper plot in Fig.~\ref{fig:BP2}, where $\delta m_h$ is very large, the electroweak potential becomes metastable and even short-lived. So, the constraints from charge and colour breaking minima are actually more severe than the ones from flavour observables.
\begin{figure}[bh]
\begin{center}
\includegraphics[width=1.\linewidth]{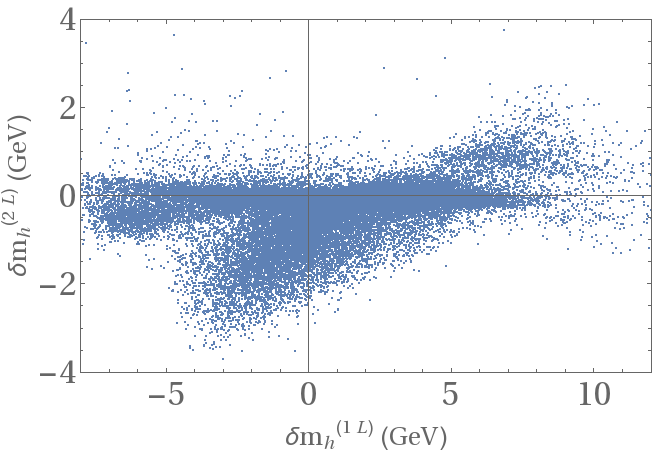}  
\end{center}
\caption{Correlation between the off-diagonal-flavour induced shift in the Higgs mass at one and two loops. }
\label{fig:OneLvsTwoL}
\end{figure}

\section{Discussion}
\label{sec:discussion}
\begin{figure}[th]
\begin{center}
\includegraphics[width=1.\linewidth]{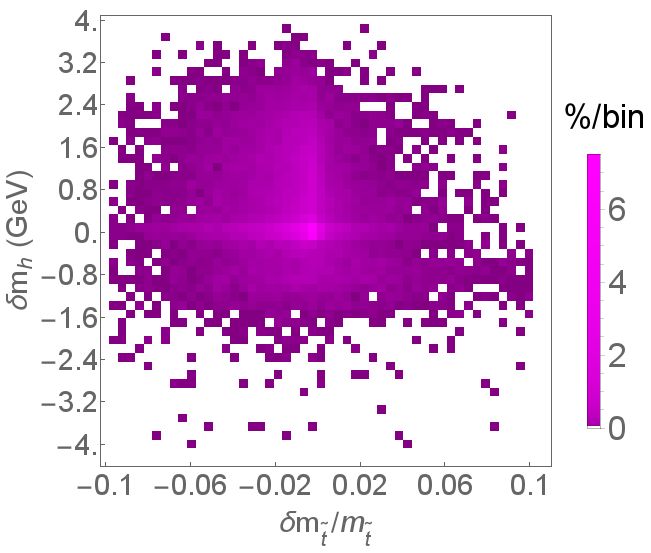}  \\[8mm]
\includegraphics[width=1.\linewidth]{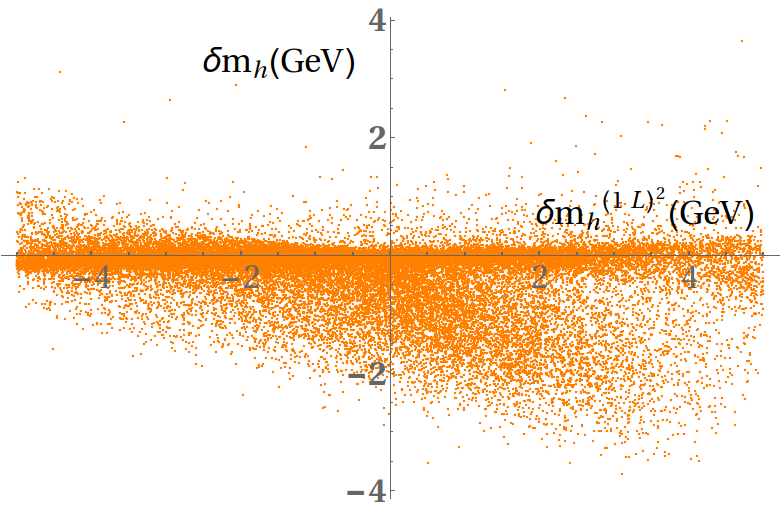}  
\end{center}
\caption{{\it Top:} $\delta m_h$ against proportional shift in lightest stop mass, $\delta m_{\tilde t_1}/m_{\tilde t_1}$ compared to model with $T_{23} = T_{32} = 0$, colours show percentage of points in each bin in a 50 by 50 grid, bins with zero points shown as white. {\it Bottom: } $\delta m_h $ (ordinate) against approximation for shift from inserting on-shell stop masses into the one-loop Higgs mass expression (abscissa) as given in equation (\ref{eq:guessapprox}). }
\label{fig:approxes}
\end{figure}

We have analysed the effect of large flavour-mixing on the Higgs mass calculation, and compared it to the approximation that only the third generation contributes, finding that the discrepancy can be several GeV for parameter points that are consistent with all other observations. The size and the sign of the flavoured two-loop contributions depends mainly on the hierarchy in the soft-breaking squark masses, the size of the flavour violating trilinear soft-terms and the gluino mass. This raises several questions:
\begin{enumerate}
\item {\it Do the shifts at two loops correlate with those at one loop?} At the beginning of section \ref{sec:numerics} we saw that there is a relationship between the shifts at one and two-loops; clearly models with large generation mixing will show large effects at one loop. We show this again for the points of our fine scan in Fig.~\ref{fig:OneLvsTwoL}. We see roughly two branches of points: the first which exhibit negligible differences between the third-generation-only approximation and full calculation at two loops, and those for which there is a positive correlation for $\delta m_h$ with the one-loop shift; i.e. broadly speaking, for the points that show devations from the third-generation approximation, the flavour-dependent corrections at one and two loops are correlated. A naive assumption for the origin of the branches would be that the gluino mass would suppress the differences at two-loops, and so the correlated points should be those with heavy gluinos. In fact, this is not the case: only the points on the correlated branch tend to exhibit large ratios of gluino to stop masses, indicating that the gluinos have the effect of enhancing the two-loop corrections in general. 
\item {\it Are the corrections proportional to the full Yukawa ($y_{u,d,c,s}\neq 0$) couplings?} To investigate this, we recalculated the corrections with only the top/bottom mass terms in the Yukawa couplings non-zero, and found very little difference; in fact only $y_t\equiv Y_u^{33}$ is relevant.Therefore, it is purely the trilinear couplings $T_{u,ij}$ that are responsible for the shifts.
\item {\it Are the differences mostly in $\alpha_t^2$ or $\alpha_t \alpha_s$ corrections?} By comparing results using specially modified versions of our code we have compared the ``full'' and third-generation only results for the strong corrections only, and found that as usual the strong corrections are largest and thus exhibit the largest differences. 
\item {\it Can the differences be explained by the effect of the one-loop shift in the stop masses?}
The one-loop shift to the light stop mass stemming from flavour terms, $\delta m_{\tilde t}=m_{\tilde t}^{1L}-m_{\tilde t}^{1L}\big|_{T_{23}=T_{32}=0}$, could be correlated to $\delta m_h$. Fig.~\ref{fig:approxes} (top panel) shows a 2D histogram of points with respect to $\delta m_{\tilde t}/m_{\tilde t}$ and $\delta m_h$. There is no clear sign of a correlation between the two, but rather a spread of points and with many large shifts in the Higgs mass showing no change in the lightest stop mass. However, as a slightly more refined measure, we could use a guess for the order of magnitude of two-loop corrections ($\delta^{\rm os} m_h^2$) as inserting the one-loop corrected stop masses into the expression for the one-loop Higgs mass: 
\begin{equation}
\delta^{\rm os} m_h^2 =  \delta^{1L} m_h^2 (m_{\tilde{t}_i}^{\rm on-shell}) - \delta^{1L} m_h^2 (m_{\tilde{t}_i}^{\DRbarp}) ,
\end{equation}
with  $\delta^{1L} m_h^2 (M)$ being the one-loop correction to the Higgs mass-squared computed using the effective potential method found, for example, in \cite{Degrassi:2001yf} (and we defined the ``stops'' as being the two eigenstates with largest stop components). Here we use the $\overline{DR}'$ values for all other parameters (mixing angles, trilinears, etc). The value that we obtain, $\delta^{\text{os}} m_h^2$, is not a true two-loop value, but merely an (in general large over-) estimate of the order of magnitude.
 However we can use this to see whether shifting both the stop masses may correlate with the Higgs mass shift we calculate, using the quantities
\begin{align}\label{eq:guessapprox}
\hspace{1cm} 
&\delta^{\rm (1L)^2} m_h^2 \equiv  \delta^{\rm os} m_h^2  - \delta^{\rm os} m_h^2\big|_{T_{32} = T_{23} = 0} \,,\\
\rightarrow& \delta m_h^{\rm (1L)^2} \equiv \bigg( (m_h^{\mathrm{approx}})^2 + \delta^{\rm (1L)^2} m_h^2\bigg)^{1/2} - m_h^{\mathrm{approx}}, \nonumber
\end{align}
where $\delta m_h^{\rm (1L)^2}$ is derived from $\delta^{\rm (1L)^2} m_h^2$ as a parameter of mass dimension 1.
The size of $\delta m_h^{\rm (1L)^2}$ gives the would-be ``two-loop'' shift in the Higgs mass when we turn on the generation-mixing trilinear using the above reasoning; we subtract off the equivalent contribution with those trilinears turned off as that is supposedly accounted for in $m_h^{\mathrm{approx}} $.  
As we show in Fig.\ref{fig:approxes} there is a weak anti-correlation between this $\delta m_h^{\rm (1L)^2}$ and $\delta m_h$, the full shift that we find. In fact, the plot appears to give the inverse of the relationship shown in Fig.\ref{fig:OneLvsTwoL}. Since this is a weak anti-correlation, it  merely reflects the relationship between the one- and two-loop shifts. However, it does imply that the discrepancy between our full two-loop calculation and the third-generation-only approximation (for the two loop parts)  might be reduced by passing from the $\DRbarp$-scheme to on-shell scheme for (at least) the stop masses. 
While this is a complicated undertaking -- requiring an on-shell scheme for at least two generations, and the inclusion of the new counterterms at two loops -- it would be interesting to explore this in future work. However, this would at best explain part of the difference: the plots show that a significant proportion of the points show no correlation at all, including some of the points with the largest differences which have almost no shift in the stop masses whatsoever. 
\end{enumerate}

From considering the above, we conclude that a sizeable contribution to $\delta m_h$ arises from the new diagrams involving the trilinear couplings $T_{23}, T_{32}$ mixing the generations, and that the effects can not be simply obtained from the existing approximate expressions. Hence, once these trilinear terms have magnitude comparable to the other soft terms we can no longer trust the third-generation-only approximation.

It would be interesting to consider consistent models realising such substantial flavour violation terms from a top-down perspective (along the lines of e.g. \cite{Galon:2013jba}). Moreover, since the result is largely independent of the charm mass, models mixing the stop and sup should yield very similar results.

\section*{Acknowledgements}
M.~D.~G thanks Pietro Slavich and Robert Ziegler for helpful discussions. This work was supported by ANR grant HIGGSAUTOMATOR (ANR-15-CE31-0002) and the Institut Lagrange de Paris. This work has been partially supported by the DFG Forschergruppe FOR 2239/1 ``New physics at the LHC''.

\bibliographystyle{JHEP-2}
\bibliography{lit}

\end{document}